\documentclass{JHEP3} 

\usepackage{epsfig,multicol}
\usepackage{amsmath, amssymb}
\usepackage{concmath, palatino}

\newcommand\fverb{\setbox\pippobox=\hbox\bgroup\verb}
\newcommand\fverbdo{\egroup\medskip\noindent%
			\fbox{\unhbox\pippobox}\ }
\newcommand\fverbit{\egroup\item[\fbox{\unhbox\pippobox}]}
\newbox\pippobox

\newcommand{\be}{\begin{equation}}
\newcommand{\ee}{\end{equation}}
\newcommand{\ba}{\begin{eqnarray}}
\newcommand{\ea}{\end{eqnarray}}

\newcommand{\refeq}[1]{Eq.~(\ref{eq:#1})}

\newcommand{\ads}{AdS_5\times S^5}

\title{On the wrapping correction to single magnon energy in twisted ${\cal N}=4$ SYM}

\author{Matteo Beccaria and Gian Fabrizio De Angelis\\
  Physics Department, Salento University, 
  Via Arnesano, 73100 Lecce\\
  INFN, Sezione di Lecce\\
  E-mail: \email{matteo.beccaria@le.infn.it} \\
}

\abstract{
We apply Zeilberger summation to  derive a closed formula for the wrapping correction to one-impurity states in the $\mathfrak{su}(2)$
sector of the $\beta$-deformed ${\cal N}=4$ SYM theory at $\beta=1/2$. As an application depending heavily on the result, 
we compute the large volume expansion of the wrapping correction.
}

\begin{document} 

\section{Introduction and result}

A major outcome of AdS/CFT duality~\cite{Maldacena:1997re} is the non perturbative integrability structure of the four dimensional planar
${\cal N}=4$ SYM theory. As a consequence, the calculation of higher-loop anomalous dimensions of single trace gauge invariant
composite operators can be performed starting from the asymptotic long-range Bethe Ansatz equations~\cite{ABA}. In the generalized spin chain 
interpretation~\cite{Thesis}, 
the long-range equations reproduce perturbation theory by means of a sequence of integrable Hamiltonians whose interaction range increases 
with the loop order~\cite{BDS}. Thus, since every operator is associated with a specific chain length, at a certain loop order  wrapping effects 
enter the calculation. Since they are not included in the asymptotic Bethe Ansatz, their computation is a mandatory ingredient in the
determination of the spectrum.

In the past years, an intense activity has been devoted to a complete determination of the leading order wrapping effects for various 
so-called twist operators~\cite{Twist}. This has been feasible by means of generalized L\"uscher finite size corrections~\cite{Luscher} made suitable for 
the AdS/CFT case. Besides, very recently, more complete treatments  have been proposed 
which in principle can go beyond the leading order and cover the full  $\mathfrak{psu}(2,2|4)$ invariant theory~\cite{TBA}.

Moving away from the maximally symmetric ${\cal N}=4$ SYM theory, it would be interesting to extend wrapping calculations
to deformed models with less symmetry, but unbroken all-loop integrability. An important theoretical laboratory of this kind 
is the superconformal ${\cal N}=1$ $\beta$-deformation of ${\cal N}=4$ SYM at real $\beta$~\cite{Leigh:1995ep}. This deformed theory is dual to
string propagation on the Lunin-Maldacena background~\cite{Lunin:2005jy}. Integrability at all-orders is known to be preserved by the 
deformation~\cite{IntDef,Beisert:2005if}.

The simplest state that can be analyzed in this theory is the length-$L$ one-impurity operator in the $\mathfrak{su}(2)_\beta$ sector.
It takes the simple form 
\be
{\cal O}_L = \mbox{Tr}\big(\varphi\,Z^{L-1}\big),
\ee
where $\varphi$ and $Z$ are the two elementary scalars appearing in the $\mathfrak{su}(2)_\beta$ composite operators.  
For $L>2$ the anomalous dimension of this operator is not protected by supersymmetry and reads~\cite{Leigh:1995ep} 
\ba
\gamma^\beta_L &=& \gamma^\beta_{L, \rm asymp} + \gamma^\beta_{L, \rm wrap}, \\
\gamma^\beta_{L, \rm asymp} &=& -1+\sqrt{1+16\,g^2\,\sin^2(\pi\,\beta)}, \\ 
\gamma^\beta_{L, \rm wrap} &=& g^{2\,L}\, \delta\gamma^\beta_L + \cdots, 
\ea
where $g^2 = \frac{g^2_{\rm YM}\,N_c}{16\pi^2}$ is the planar coupling. The piece $g^{2\,L}\,\delta\gamma^\beta_L$ is the leading order wrapping correction
which has been discussed in great details in \cite{Fiamberti:2008sm,Fiamberti:2008sn}. By direct diagram calculations in superspace, it is possible
to compute $\delta\gamma^\beta_L$ for each given $L$ by solving a recursion relation between  multi-loop basic diagrams. This construction is 
carried over up to $L=9$ revealing a very clean transcendentality structure of the result. 

A major simplification occurs when $\beta=1/2$ and $L$ is even. This is in a sense the most simple twisting of ${\cal N}=4$. 
Indeed, the Bethe Ansatz equations are the same as in the undeformed case $\beta=0$ apart from a change in the cyclicity constraint
for single trace operators. As a consequence, following the analysis of~\cite{Gomez:2008hx,Gunnesson:2009nn}, 
the wrapping correction $\delta\gamma^{1/2}_L$ can be computed  in the undeformed theory with relaxed cyclicity constraint 
in terms of the wrapping correction to a single magnon in $\mathfrak{sl}(2)$ sector with unphysical momentum $p=\pi$.
This is a calculation that can be done with the techniques of~\cite{Twist}. The result is an efficient algebraic algorithm for the 
calculation of  $\delta\gamma^{1/2}_L$ which provides a quick answer for each $L$. This method agrees in all available cases, 
with the superspace calculation and has the advantage of bypassing the recursive evaluation of multi-loop diagrams.

Nevertheless, apart from the maximal transcendentality term in the answer, we still do not get a closed formula for the correction.
The aim of this brief paper is that of providing such formula. It is very simple and compact and reads ($L$ even)
\be
\label{eq:result}
\delta\gamma^{1/2}_L = -64\,\binom{2\,L-3}{L-1}\,\zeta(2\,L-3) + 128\,\sum_{\ell=1}^{\frac{L}{2}-1}
\frac{2^{2\,\ell}\,\ell}{L-2\,\ell-1}\,\binom{2\,L-2\,\ell-3}{L-1}\,\zeta(2\,L-2\,\ell-3),
\ee
Specific cases are reported in Appendix B. As a simple application of \refeq{result}, we shall compute the large $L$ behavior of the correction. This is a 
quantity which crucially needs the closed result \refeq{result}. The leading term in the large volume expansion turns out to be 
$\delta\gamma^{1/2}_L \sim 4^L\,L^{-3/2}$ giving an exponentially vanishing {\em critical} wrapping correction if $g^2 < 1/4$. Notice that the exponential term 
is shared by the maximal transcendentality contribution, but the algebraic correcting factor is different due to various cancellations in \refeq{result}.

In our opinion, the derivation of \refeq{result} is not trivial. In particular, we needed some sophisticated summation algorithms
that, we believe, could deserve a certain interest by themselves. For this reason, we have tried to present the necessary tools in a self-contained way with some
illustrative simple example. 

\section{The wrapping correction from superspace diagrams}

As we recalled in the Introduction, the main result of~\cite{Fiamberti:2008sn} is the  
first wrapping correction $\delta\gamma^\beta_L\,g^{2L}$ to the anomalous dimension of the 1-impurity operator 
$\mbox{Tr}(\varphi\,Z^{L-1})$ in the $\mathfrak{su}(2)$ sector of the $\beta$-deformed $N=4$ SYM theory. Here, we shall be interested in the $\beta=1/2$
case where the undeformed theory is twisted in the simplest way. We shall also restrict our analysis to the case of 
even $L$ for technical reasons which will be clearer later. For $\beta=1/2$ and even $L$ the wrapping correction reads
\ba
\label{eq:zanon}
\delta\gamma^{1/2}_L &=& -32\,L\,\left(P_L-2\,\sum_{j=0}^{\left[\frac{L}{2}\right]-1}(-1)^j\,I_L^{(j+1)}\right), \\
P_L &=& \frac{2}{L}\binom{2L-3}{L-1}\,\zeta(2L-3),
\ea
where $I_L^j$ can be recursively extracted from multi-loop diagrams and are explicitly tabulated up to $L=9$ in~\cite{Fiamberti:2008sn}. 
For the present discussion it is important to remark that no general closed expression is available for these crucial quantities.
In other words, the quantity $\delta\gamma^{1/2}_L$ can be in principle be computed for any given $L$, but not as a closed function 
of this physically meaningful parameter.

From inspection of the first cases, $L=4, 6, 8, \dots$, one is led to {\em conjecture} the following general form 
of the wrapping correction
\be
\label{eq:Ansatz}
\delta\gamma^{1/2}_L = \sum_{\ell=0}^{\frac{L}{2}-1} a_{L, \ell}\,\zeta(2L-3-2\,\ell),
\ee
with {\em integer} coefficients $a_{L, \ell}$. Also, the maximum transcendentality term comes entirely from the $P_L$ term in \refeq{zanon}. 
This means that one coefficient is known
\be
\label{eq:maximal}
a_{L, 0} = -64\,\binom{2L-3}{L-1}.
\ee

\section{The wrapping correction from unphysical undeformed spectrum}

In the approach pursued in~\cite{Gunnesson:2009nn,Gomez:2008hx}, the wrapping correction of \refeq{zanon} can be identified with
the wrapping correction to the energy of a single unphysical magnon with momentum $p=\pi$ in the undeformed theory. 
This can be computed by L\"uscher-Janik finite size corrections in the $\mathfrak{sl}(2)$ sector.

The resulting formula is quite simple~\footnote{The factor $8^2$ is the square of the one magnon one-loop 
anomalous dimension} 
\be
\delta\gamma^{1/2}_L = -8^2\cdot 4^{L-2}\,\sum_{Q=1}^\infty\int_{-\infty}^\infty\frac{dq}{2\pi}\frac{T^2(q,Q)}{R(q,Q)}\,\frac{1}{(q^2+Q^2)^{L-2}},
\ee
where $T(q, Q)$ and $R(q, Q)$ are polynomials in $q$ and $Q$ which can be expressed in terms of the 
(almost trivial) Baxter function associated with the single magnon state with $p=\pi$. Their general expression can be 
found in~\cite{Gunnesson:2009nn} and will not be needed here.
Kinematical arguments show that, under summation over $Q$, the integral over the rapidity $q$ is given by the residue at the pole $q=i\,Q$.
Thus we can write
\ba
\delta\gamma^{1/2}_L &=& 
-4^{L+1}\,i\,\sum_{Q=1}^\infty\mathop{\mbox{Res}}_{q=i\,Q} \left(
\frac{T^2(q,Q)}{R(q,Q)}\,\frac{1}{(q^2+Q^2)^{L-2}}
\right) = \nonumber \\
&=&
-4^{L+1}\,i\,\sum_{Q=1}^\infty\mathop{\mbox{Res}}_{q=i\,Q} \Lambda_L(q, Q),
\ea
where $\Lambda_L(q, Q)$ is 
\be
\Lambda_L(q, Q) = \frac{T^2(q,Q)}{R(q,Q)}\,\frac{1}{(q^2+Q^2)^{L-2}}.
\ee
Replacing the explicit expressions of $T$, and $R$ we have
\be
\Lambda_L(q, Q) = \frac{4\,Q^2\,(q^2+Q^2-1)^2}{(q^2+Q^2)^L\,\left[q^2+(Q+1)^2\right]\,\left[q^2+(Q-1)^2\right]}.
\ee
On general grounds we can easily prove that 
\be
\mathop{\mbox{Res}}_{q=i\,Q} \, \Lambda_L(q, Q) = \frac{P_L(Q)}{Q^{2L-3}(4Q^2-1)^L},
\ee
where $P_L(Q)$ is an even polynomial of degree $\deg P_L = 3L-2$.
Again, one can inspect the first cases, $L=4, 6, 8, \dots$. In all instances, the residue can be written as
\be
\label{eq:tmp1}
\frac{P_L(Q)}{Q^{2L-3}(4Q^2-1)^L} = \sum_{\ell=0}^{\frac{L}{2}-1}\frac{a_{L,\ell}'}{Q^{2L-3-2\,\ell}} + R_L(Q),
\ee
where $R_L(Q)$ is a rational function regular in $Q=0$ and such that 
\be
\label{eq:tmp2}
\sum_{Q=1}^\infty R_L(Q) = 0.
\ee
The explicit value of the coefficients of the polar part obeys $a_{L,\ell} = a'_{L,\ell}$ and  matches perfectly those in \refeq{Ansatz}

\section{Proof of the closed formula}

Our aim, will be that of establishing \refeq{Ansatz} rigorously with a simple closed formula for $a_{L, \ell}$. To this aim, we shall
pursue the residue formula described in the previous section and perform the following 3 steps:
\begin{enumerate}
\item  We compute the coefficients of the Laurent expansion of $\mathop{\mbox{Res}}_{q=i\,Q} \, \Lambda_L(q, Q)$ in $Q=0$
as complicated finite sums involving ${\cal O}(L)$ terms.
\item We find a closed form of the poles coefficients by application of the Zeilberger's summation algorithm.
For the polar part, these coefficients are clearly in one to one relation with the coefficients $a_{L,\ell}$ in \refeq{Ansatz}. 
\item We prove that the rational function $R_L$ appearing in \refeq{tmp1} obeys indeed \refeq{tmp2}.
\end{enumerate}

\subsection{Step 1: Implicit formula for the Laurent coefficients}

We start from the trivial decomposition
\be
\label{eq:decomp}
\Lambda_L(q, Q) = \frac{4\,Q^2}{(q^2+Q^2)^L} +\widetilde\Lambda_L(q, Q)+\widetilde\Lambda_L(q, -Q),
\ee
where
\be
\label{eq:lambdatilde}
\widetilde\Lambda_L(q, Q) = -\frac{4\,Q\,(Q+1)^2}{q^2+(Q+1)^2}\,\frac{1}{(q^2+Q^2)^L}.
\ee
The first term in \refeq{decomp} is responsible for the maximal transcendentality contribution as discussed in~\cite{Gunnesson:2009nn}. 
Its contribution is known, see \refeq{maximal}, and we focus on the 
other terms. We set $q=i\,Q+z\,Q$ and take the Laurent expansion of \refeq{lambdatilde} around $z=0$
\be
\widetilde\Lambda(i\,Q+z\,Q, Q) = \frac{-4\,(Q+1)^2}{Q^{2L-1}\,\omega^L\,(1+2\,Q+Q^2\,\omega)},
\ee
where
\be
\omega = z^2+2\,i\,z = 2\,i\,z\,\left(1-i\,\frac{z}{2}\right).
\ee
Expanding first in powers of $\omega$ and then in powers of $z$ we find 
\be
\widetilde\Lambda(i\,Q+z\,Q, Q) = -\frac{4\,(Q+1)^2}{Q^{2L-1}}\sum_{k\ge 0}(-1)^k\,\frac{Q^{2k}}{(1+2Q)^{k+1}}\,(2\,i\,z)^{k-L}
\sum_{p\ge 0}\binom{k-L}{p}\,\left(\frac{z}{2\,i}\right)^p.
\ee
We can now extract the coefficient $\widetilde{c}_L$ of the simple pole $\widetilde\Lambda(i\,Q+z\,Q, Q)  = \cdots + \widetilde{c}_L/z + \cdots$. It is 
$Q$ times the residue in the initial $q$ variable. It reads
\be
\widetilde{c}_L = -\frac{8\,i\,(Q+1)^2}{Q^{2L-1}}\,2^{-2L}\,\sum_{k\ge 0} 2^{2\,k}\,\frac{Q^{2\,k}}{(1+2\,Q)^{k+1}}\,\binom{k-L}{L-k-1} .
\ee
Expanding the powers of $1/(1+2Q)$, we find
\be
\widetilde{c}_L = -\frac{i\,2^{3-2L}(Q+1)^2}{Q^{2L-1}}\,\sum_{k\ge 0}\sum_{p\ge 0}  (2\,Q)^{2\,k+p}\,\binom{k-L}{L-k-1}\,\binom{-k-1}{p} .
\ee
The sum can be reorganized as
\ba
\widetilde{c}_L &=& -i\,2^{4-2L}\,\sum_{\ell=1}^\infty \frac{2^{2\,\ell}}{Q^{2L-2\ell-2}}\sum_{k=0}^\ell \binom{k-L}{L-k-1}\times\\
&& \times \,\left[
\binom{-k-1}{2\ell-2k+1} + \binom{-k-1}{2\ell-2k}+\frac{1}{4}\,\binom{-k-1}{2\ell-2k-1} 
\right] = \\
&=&  -i\,2^{4-2L}\,\sum_{\ell=1}^\infty \frac{2^{2\,\ell}}{Q^{2L-2\ell-2}}\,h_{L, \ell},
\ea
where
\be
\label{eq:h}
h_{L, \ell} = \sum_{k=0}^\ell\binom{k-L}{L-k-1}\binom{-k-1}{2\ell-2k}\frac{k-\ell-2\ell^2}{2\,(2\ell-k)\,(2\ell-2k+1)}.
\ee
This is a rather complicated finite sum. In practice, to obtain a closed formula for the wrapping correction
is completely equivalent to finding the same for this sum. Indeed, in this section, we have just shown that 
\be
\delta\gamma^{1/2}_L = \sum_{Q=1}^\infty \delta\gamma^{1/2}_L(Q),
\ee
with
\be
\delta\gamma_L^{1/2}(Q) = -64\,\binom{2L-3}{L-1}\frac{1}{Q^{2L-3}}-128\,\sum_{\ell=1}^\infty
\frac{2^{2\ell}}{Q^{2L-2\ell-3}}\,h_{L, \ell}.
\ee

\subsection{Step 2: Performing the finite summation}

Up to now, we have just reshuffled the L\"uscher-Janik's correction formula.
The novelty comes when one tries to perform in closed form the finite sum defining 
$h_{L, \ell}$, namely \refeq{h}. This can be done by applying a powerful tool, the Zeilberger's algorithm~\cite{Zeil}.
It is a very nice mathematical device to perform rather difficult finite summations. We now briefly describe it referring 
to~\cite{book} for a more detailed discussion. Then, we present its application to \refeq{h}.

\subsubsection{Algorithms for hypergeometric summation}

Let us consider the finite sum
\be
\label{eq:problem}
S_L = \sum_{k\in\mathbb{Z}} \sigma_{L, k}, 
\ee
where $\sigma_{L,k}\neq 0$ in a $L$-dependent finite interval $k_{\rm min}(L) \le k \le k_{\rm max}(L)$. Is it possible to write 
the sum in closed form as a function of $L$ ? A general strategy to face this problem is available when $\sigma_{L,k}$ is 
doubly hypergeometric, {\em i.e.} the ratios $\sigma_{L+1, k}/\sigma_{L,k}$ and $\sigma_{L, k+1}/\sigma_{L,k}$ are rational functions of $L$ and $k$.
The strategy is known as Sister Celine's algorithm and consists in finding positive integers $I$, $J$ such that 
\be
\label{eq:celine}
\sum_{i=0}^I\sum_{j=0}^J c_{i,j}(L)\,\sigma_{L+i, k+j} = 0.
\ee
Basically, one divides by $\sigma_{L, k}$, simplifies, put everything under a common denominator,  and equates to zero the 
various powers of $k$ in the numerator. Under very mild conditions, such a doubly recursive relation can always be found if $I$ and $J$ are taken 
large enough. Summing over $k$, we derive a recursion for the finite sum
\be
\sum_{i=0}^I \left(\sum_{j=0}^J c_{i,j}(L)\right)\,S_{L+i} = 0.
\ee
This can be used to find a closed form for \refeq{problem} or, possibly, to show that a closed form does not exists in a given class~\footnote{
An example is the class of finite linear combination of hypergeometric terms, fully treated in~\cite{Pet}.}

Zeilberger's summation algorithm is a very efficient version of Sister Celine's one. It provides an instance of \refeq{celine} in the nice form
\be
\label{eq:zeil}
\mathbb{D}_L \sigma_{L, k} = \Delta_k (\sigma_{L, k}\,R_{L, k}),  \\
\ee
where
\ba
\mathbb{D}_L f_L &=& \sum_{i=0}^I c_{i}(L)\,f_{L+i}, \\
\Delta_k f_k &=& f_{k+1}-f_k, 
\ea
and $R_{L, k}$ is a rational function. By summing over $k$, this relation immediately provides the desired recursion relation 
\be
\mathbb{D}_L S_L = 0,
\ee
since the right hand side telescopes. Explicit illustrative examples are collected in Appendix A.

\subsubsection{Application to $h_{L,\ell}$}

In order to apply Zeilberger's algorithm it is necessary to modify a little the presentation of $h_{L, k}$. To this aim, we 
use the following identity valid for the usual extension of the binomial coefficient
\be
\binom{a}{b} = (-1)^b\,\binom{b-a-1}{b},
\ee
and write
\ba
h_{L, \ell} = \sum_{k=0}^\ell \sigma_{L, \ell, k},
\ea
with
\be
\label{eq:xxx}
\sigma_{L, \ell, k} = -(-1)^k\,\binom{2L-2k-2}{L-k-1}\,\binom{2\ell-k}{2\ell-2k}\,\frac{k-\ell-2\ell^2}{2(2\ell-k)(2\ell+1-2k)}.
\ee
A double application of Zeilberger's algorithm with respect to $L$ or $\ell$ provides the following two recursions
\ba
\label{eq:rec1}
&& (1 + \ell)\, (1 + 2\, \ell - L)\, (2 + 2\, \ell - L)\, \sigma_{L, \ell, k} \\
&& \qquad -  2\, \ell \, (3 + 2 \, \ell - 2\, L) \,(2 + \ell - L) \, \sigma_{L, \ell+1, k} = \Delta_k(\sigma_{L, \ell, k}\, F_{L, \ell, k}), \nonumber \\
\label{eq:rec2}
&& -2 \, (2\, \ell-2\, L+1) \,(\ell-L+1) \,\sigma_{L, \ell, k}-(2\, \ell-L)\, L\, \sigma_{L+1, \ell, j} =  \Delta_k(\sigma_{L, \ell, k} \,G_{L, \ell, k}),
\ea
with 
\ba
F_{L, \ell, k} &=& \frac{k \,(2\, \ell-k)\, (-2\, k+2\, L-1)}{2\, (-k+\ell+1) \,(-2 \, k+2\, \ell+3) \left(2\, \ell^2+\ell-k\right)}\,\left[
2\, (3\, \ell-2\, L+4)\, k^2+\right.\\
&&\left. + 2\, \left(6\, \ell^3+(3-4\, L)\, \ell^2+(3\, L-14)\, \ell+5\, L-10\right)\, k + \right. \nonumber \\
&& \left. -2\, \left(8\, \ell^4+(18-6\, L) \,\ell^3+(3-5\, L)\,
   \ell^2+(4\, L-13)\, \ell+3\, L-6\right)
\right],\nonumber \\
G_{L, \ell, k} &=& \frac{k \,(2 \,\ell-k)\, \left(2\, k+2\, \left(2 \,\ell^2+l-2\, L\right)\right)\, (-2\, k+2\, L-1)}{\left(2\, \ell^2+l-k\right)\, (L-k)}.
\ea
As we remarked, \refeq{rec1} and \refeq{rec2} can be checked by direct substitution of \refeq{xxx}. Summing over $k$ we find the two recursions
for the sum $h_{L, \ell}$
\ba
\label{eq:rec12}
&& (1 + \ell)\, (1 + 2\, \ell - L)\, (2 + 2\, \ell - L)\, h_{L, \ell} + \nonumber \\
&& \qquad -  2\, \ell \, (3 + 2 \, \ell - 2\, L) \,(2 + \ell - L) \, h_{L, \ell+1} = 0, \\
\label{eq:rec22}
&& -2 \, (2\, \ell-2\, L+1) \,(\ell-L+1) \,h_{L, \ell}-(2\, \ell-L)\, L\, h_{L+1, \ell} =  0.
\ea
In a subset $\Omega$ of the discrete  $(L, \ell)$ plane which is maximally connected by the recursion relations, the only solution 
to \refeq{rec12} and \refeq{rec22} is 
\be
h_{L, \ell} = {\cal C}_\Omega\,\frac{\ell}{L-2\,\ell-1}\,\binom{2L-2\ell-3}{L-1},
\ee
where ${\cal C}_\Omega$ is a constant. To prove this statement it is enough to define
\be
\widehat{h}_{L, \ell} = h_{L, \ell}\,\left[\frac{\ell}{L-2\,\ell-1}\,\binom{2L-2\ell-3}{L-1}\right]^{-1}, 
\ee
and check that the two recursions simplify to 
\ba
&& \widehat h_{L, \ell} - \widehat h_{L, \ell+1} = 0, \\
&& \widehat h_{L, \ell} - \widehat h_{L+1, \ell} =  0,
\ea
showing that $\widehat{h}_{L, \ell}$ is constant over any connected region.

\subsubsection{Closed form of the Laurent expansion}

We are interested in the cases $\ell\ge 1$. It follows that there are two disconnected regions $\Omega$.
The first is $1\le \ell \le\frac{L}{2}-1$ and is associated with the polar part of the Laurent expansion. The second is 
$\ell\ge L-1$ and defines an infinite power series, regular around $Q=0$. Fixing ${\cal C}_\Omega$ by means of one special 
value in each region, we find the final result for $\delta\gamma^{1/2}_L(Q)$
\ba
\delta\gamma^{1/2}_L(Q) &=& \mbox{pol}_L(Q) + \mbox{reg}_L(Q), \nonumber \\
\mbox{pol}_L(Q) &=&  -64\,\binom{2\,L-3}{L-1}\frac{1}{Q^{2\,L-3}} + 128\,\sum_{\ell=1}^{\frac{L}{2}-1}
\frac{2^{2\,\ell}\,\ell}{L-2\,\ell-1}\,\binom{2\,L-2\,\ell-3}{L-1}\,\frac{1}{Q^{2\,L-2\,\ell-3}}, \nonumber\\
\label{eq:reg}
\mbox{reg}_L(Q) &=& 16\,2^{2\,L}\,\sum_{\ell=0}^\infty 2^{2\,\ell}\,\frac{L-1+\ell}{L-1+2\,\ell}\,\binom{L+2\,\ell-1}{L-1}\,Q^{2\,\ell+1}.
\ea
After sum over $Q$, the polar part gives precisely the result \refeq{result}. To finish, we have just to show that the regular part does not
give any contribution. This is shown in the next section.

\subsection{Step 3: Vanishing of the rational part}

We want to prove that 
\be
\sum_{Q=1}^\infty \mbox{reg}_L(Q) = 0.
\ee
This follows from the remarkable identity
\be
\label{eq:idrat}
\mbox{reg}_L(Q)  = 2^{2L+3}\,Q\,\left[\frac{Q+1}{(2\,Q+1)^L}-\frac{Q-1}{(2\,Q-1)^L}\right].
\ee
\refeq{idrat} is easily proved since (a) the Taylor expansion around $Q=0$ of the right hand side coincides with the 
regular expansion in \refeq{reg}, and (b) we know a priori that $\mbox{reg}_L(Q)$ is a rational function of $Q$. \refeq{idrat}
follows from (a)+(b) and 
the fact that a rational function is completely determined by its Laurent expansion around any point. Taking the sum over $Q$, 
we can write
\be
\sum_{Q=1}^\infty \mbox{reg}_L(Q) = \sum_{Q=1}^\infty [F(Q+1)-F(Q)], \qquad F(Q) =  2^{2L+3}\,\frac{Q\,(Q-1)}{(2\,Q-1)^L},
\ee
which clearly vanishes for the physically relevant cases $L>2$.

\section{An application of the closed formula: The large $L$ expansion}

As an application of \refeq{result}, let us derive the large $L$ asymptotics of the critical wrapping correction. At large $L$ we can definitely replace the $\zeta$ functions
by 1 up to exponentially small terms. We thus have to compute the large $L$ expansion of the two pieces
\ba 
A_L &=& -64\,\binom{2\,L-3}{L-1}, \\
B_L &=& 128\,\sum_{\ell=1}^{\frac{L}{2}-1} \frac{2^{2\,\ell}\,\ell}{L-2\,\ell-1}\,\binom{2\,L-2\,\ell-3}{L-1}.
\ea
The expansion of the first term is trivial, by Stirling large $L$ expansion
\be
A_L = \frac{4^L}{\sqrt\pi}\,\left(-\frac{8}{L^{1/2}}-\frac{3}{L^{3/2}}-\frac{25}{16}\,\frac{1}{L^{5/2}} + \cdots\right).
\ee
The second term is more involved. Another application of Zeilberger's algorithm puts it in the nicer form 
\be
B_L = 8\,\left(\frac{16}{9}\right)^{L/2-1}\,\sum_{k=0}^{\frac{L}{2}-2}\,\left(\frac{9}{16}\right)^k\,\frac{8\,k+9}{k+1}\,\binom{4\,k+4}{2\,k+1}.
\ee
Expanding the summand, and dropping the term $k=0$ which is exponentially suppressed with respect to the full sum, we have
\be
B_L = \frac{8}{\sqrt{2\,\pi}}\,\left(\frac{16}{9}\right)^{L/2-1}\,\sum_{k=1}^{\frac{L}{2}-2}\,9^k\,\left(
\frac{128}{k^{1/2}}-\frac{120}{k^{3/2}}+\frac{637}{4}\,\frac{1}{k^{5/2}} + \cdots\right).
\ee
The sums are
\be
\sum_{k=1}^{\frac{L}{2}-2} \frac{9^k}{k^s} = -9^{\frac{L}{2}-1}\,\Phi\left(9, s, \frac{L}{2}-1\right) + \mbox{Li}_s(9).
\ee
where $\Phi$ is the Lerch transcendent function
\be
\Phi(z, s, a) = \sum_{n=0}^\infty\frac{z^n}{(n+a)^s}.
\ee
The polylogarithm contribution is exponentially suppressed. Dropping it, we find 
\ba
B_L &=& \frac{1}{2\sqrt{2\,\pi}}\,4^L\,\left(
-128\,\Phi\left(9, \frac{1}{2}, \frac{L}{2}-1\right)+120\,\Phi\left(9, \frac{3}{2}, \frac{L}{2}-1\right) + \right.\nonumber\\
&&  \left. -\frac{637}{4}\,\Phi\left(9, \frac{5}{2}, \frac{L}{2}-1\right) + \cdots\right).
\ea
The large $L$ expansion of these Lerch functions can be obtained from the large $N$ expansion of its integral representation
valid for $z<1$,
\be
\Phi(z, s, N) = \frac{1}{\Gamma(s)}\int_0^\infty\frac{t^{s-1}\,e^{-N\,t}}{1-z\,e^{-t}}\,dt,
\ee
and analytically continuing to $z=9$. This gives 
\ba
\Phi\left(9, \frac{1}{2}, N\right) &=& -\frac{1}{8}\,\frac{1}{N^{1/2}}-\frac{9}{128}\,\frac{1}{N^{3/2}}-\frac{135}{2048}\,\frac{1}{N^{5/2}}  + \cdots, \\
\Phi\left(9, \frac{3}{2}, N\right) &=& -\frac{1}{8}\,\frac{1}{N^{3/2}}-\frac{27}{128}\,\frac{1}{N^{5/2}} + \cdots, \\
\Phi\left(9, \frac{5}{2}, N\right) &=& -\frac{1}{8}\,\frac{1}{N^{5/2}} + \cdots.
\ea
and therefore, 
\be
B_L = \frac{4^L}{\sqrt\pi}\,\left(+\frac{8}{L^{1/2}}+\frac{2}{L^{3/2}}+\frac{1}{16}\,\frac{1}{L^{5/2}} + \cdots\right).
\ee
Summing the $A_L$ contribution we see that the leading term cancels and we end up with 
\be
\delta\gamma^{1/2}_L = -\frac{4^L}{L^{3/2}\,\sqrt{\pi}}\,\left(1 + \frac{3}{2\,L} + \cdots\right).
\ee
This expansion is perfectly matched by the numerical values that we could compute at very large $L$ thanks to the closed formula.
As a remark, we emphasize that an estimate based on the only maximal transcendentality term would have predicted the correct exponential 
factor $4^L$, but a wrong algebraic correction $\sim L^{-1/2}$ instead of the correct one  $\sim L^{-3/2}$. 

\section*{Acknowledgments}
We thank N. Gromov and P. Vieira for useful conversations on twisted ${\cal N}=4$ SYM.

\appendix
\section{Examples of Zeilberger's summation}

As illustration we can consider the following interesting examples
\be
S^{(p)}_L = \sum_{k=0}^L (-1)^k\, \binom{L}{k}^p, \quad p = 1, 2, 3,
\ee
which can be summed as follows
\ba
S^{(1)}_L &=& 0,\\
\label{eq:sum2}
S^{(2)}_{2L} &=& (-1)^L\,2^L\,\frac{(2L-1)!!}{L!},\quad S^{(2)}_{2L+1} = 0,\\
\label{eq:sum3}
S^{(3)}_{2L} &=& (-1)^L\,\frac{(3L)!}{(L!)^3},\qquad\qquad~ S^{(3)}_{2L+1} = 0.
\ea
The first is trivial, the second easy, the third quite difficult. Let us treat them symmetrically with the Zeilberger's algorithm. 
In all cases, we extend the sum and write
\be
S^{(p)}_L = \sum_{k\in\mathbb{Z}} \sigma^{(p)}_{L, k},\qquad \sigma^{(p)}_{L, k} = (-1)^k\, \binom{L}{k}^p.
\ee

\medskip
\noindent
\underline{\bf $p=1$}
\medskip

\noindent
For $\sigma^{(1)}_{L, k}$ Zeilberger's algorithm provides the recursion
\be
\sigma^{(1)}_{L, k} = \Delta_k\left(-\frac{k}{L}\,\sigma^{(1)}_{L, k}\right).
\ee
Explicitly, this is just Pascal's relation
\be
\binom{L}{k} = \frac{k+1}{L}\binom{L}{k+1}+\frac{k}{L}\binom{L}{k} = 
\binom{L-1}{k}+\binom{L-1}{k-1}.
\ee
Multiplying by $(-1)^k$ and summing over $k$, we immediately obtain $S_L^{(1)} = 0$.

\medskip
\noindent
\underline{\bf $p=2$}
\medskip

\noindent
For $\sigma^{(2)}_{L, k}$ Zeilberger's algorithm provides the recursion
\be
(L+2)\,\sigma^{(2)}_{L+2, k} +4\,(L+1)\,\sigma^{(2)}_{L, k} = \Delta_k(\sigma^{(2)}_{L, k}\,R_{L,k}),
\ee
with 
\ba
R_{L, k} &=& -\frac{k^2}{(L-k+1)^2(L-k+2)^2}\left[2 (L+1) k^2-2 (L+1) (3 L+5) k + \right. \\
&& \left. +(L+1) \left(5 L^2+17 L+14\right)\right].\nonumber
\ea
Summing over $k$ we find the non-trivial recursion relation 
\be
(L+2)\,S^{(2)}_{L+2} +4\,(L+1)\,S^{(2)}_{L} = 0.
\ee
Using the initial values, we obtain \refeq{sum2}.

\medskip
\noindent
\underline{\bf $p=3$}
\medskip

\noindent
Finally, in the case $p=3$, Zeilberger's algorithm gives
\be
(L+2)^2\,\sigma^{(3)}_{L+2, k}+3\,(3L+4)\,(3L+2)\,\sigma^{(3)}_{L, k} = \Delta_k(\sigma^{(3)}_{L, k}\,R_{L, k}),
\ee
with
\ba
R_{L, k} &=& -\frac{k^3}{(-k+L+1)^3 (-k+L+2)^3}\,\left[
3 (3 L+4) k^4-3 (3 L+4) (5 L+8) k^3+\right. \\
&& \left. + 3 \left(29 L^3+132 L^2+198 L+98\right) k^2-3 (L+2) \left(26 L^3+109 L^2+151 L+69\right) k+ \right. \nonumber \\
&& \left. + 2 (L+2)^2 \left(14
   L^3+54 L^2+69 L+29\right)\right].\nonumber
\ea
Summing over $k$, we find the recursion 
\be
(L+2)^2\,S_{L+2}+3\,(3L+4)\,(3L+2)\,S_L = 0, 
\ee
from which the result \refeq{sum3} follows.

\section{List of wrapping corrections}

\ba
\delta\gamma^{1/2}_{4} &=& 128 \left(4 \zeta _3-5 \zeta _5\right), \nonumber\\ 
\delta\gamma^{1/2}_{6} &=& 128 \left(32 \zeta _5+28 \zeta _7-63 \zeta _9\right), \nonumber\\ 
\delta\gamma^{1/2}_{8} &=& 768 \left(32 \zeta _7+64 \zeta _9+44 \zeta _{11}-143 \zeta _{13}\right), \nonumber\\ 
\delta\gamma^{1/2}_{10} &=& 128 \left(1024 \zeta _9+3520 \zeta _{11}+4576 \zeta _{13}+2860 \zeta _{15}-12155 \zeta _{17}\right), \nonumber\\ 
\delta\gamma^{1/2}_{12} &=& 256 \left(2560 \zeta _{11}+13312 \zeta _{13}+26208 \zeta _{15}+28288 \zeta _{17}+16796 \zeta _{19}-88179 \zeta _{21}\right), \nonumber\\ 
\delta\gamma^{1/2}_{14} &=& 256 \left(12288 \zeta _{13}+89600 \zeta _{15}+243712 \zeta _{17}+372096 \zeta _{19}+361760 \zeta _{21} \right. \nonumber\\
&& \left. +208012 \zeta _{23}-1300075 \zeta _{25}\right), \nonumber\\ 
\delta\gamma^{1/2}_{16} &=& 512 \left(28672 \zeta _{15}+278528 \zeta _{17}+992256 \zeta _{19}+1984512 \zeta _{21}+2615008 \zeta _{23} \right. \nonumber\\
&& \left. +2377280 \zeta _{25}+1337220 \zeta _{27}-9694845 \zeta _{29}\right), \nonumber\\ 
\delta\gamma^{1/2}_{18} &=& 128 \left(524288 \zeta _{17}+6537216 \zeta _{19}+29417472 \zeta _{21}+73835520 \zeta _{23}+123059200 \zeta _{25} \right.\nonumber\\
&& \left. +147251520 \zeta _{27}+127743840 \zeta _{29}+70715340 \zeta _{31}-583401555 \zeta _{33}\right), \nonumber\\ 
\delta\gamma^{1/2}_{20} &=& 256 \left(1179648 \zeta _{19}+18350080 \zeta _{21}+101556224 \zeta _{23}+310886400 \zeta _{25}+631488000 \zeta _{27}\right.\nonumber\\ 
&& \left. +932305920 \zeta _{29}+1042120800 \zeta _{31}+873396480 \zeta _{33}+477638700 \zeta _{35}-4418157975 \zeta _{37}\right).\nonumber
\ea

\end{document}